\documentclass[manuscript]{aastex}
\usepackage{amsmath}
\usepackage{color}

\newcommand{\del}[1]{\delta\left(#1\right)}

\newcommand{\dvg}{\nabla\cdot}
\newcommand{\ep}{\varepsilon}
\newcommand{\dprime}{{\prime\prime}}
\newcommand{\fder}[2]{\frac{\delta #1}{\delta #2}}
\newcommand{\fint}{\int^{\infty}_{-\infty}\!\!\!}

\newcommand{\ubf}{\mathbf{u}}
\newcommand{\vb}{\mathbf{v}}
\newcommand{\G}{\mathbf{G}}
\newcommand{\Gto}{\mathbf{\tilde{G}_0}}
\newcommand{\Gt}{\mathbf{\tilde{G}}}
\newcommand{\Gtij}{\tilde{G}_{ij}}
\newcommand{\Gtkl}{\tilde{G}_{kl}}
\newcommand{\Ls}{\mathcal{L}}
\newcommand{\Lo}{\mathcal{L}_0}
\newcommand{\I}{\mathbf{I}}
\newcommand{\w}{\omega}
\newcommand{\tp}{t^{\prime}}
\newcommand{\x}{\mathbf{x}}
\newcommand{\xo}{\mathbf{x}_0}
\newcommand{\s}{\boldsymbol{\sigma}}
\newcommand{\sga}{\boldsymbol{\sigma}_a}
\newcommand{\sgb}{\boldsymbol{\sigma}_b}
\newcommand{\sgp}{\boldsymbol{\sigma}^\prime}
\newcommand{\Up}{\boldsymbol{\Upsilon}}

\newcommand{\ut}{\mathbf{\tilde{u}}}
\newcommand{\Pp}{\tilde{P}^{\prime}}
\newcommand{\g}{\mathbf{g}}
\newcommand{\go}{\mathbf{g}_0}

\newcommand{\F}{\mathbf{F}}
\newcommand{\Ft}{\mathbf{\tilde{F}}}
\newcommand{\xp}{\mathbf{x^\prime}}
\newcommand{\xpp}{\mathbf{x^\dprime}}
\newcommand{\ssi}{\int_{\partial\sun}}
\newcommand{\svi}{\int_{\sun}}
\newcommand{\q}{\mathbf{q}}
\newcommand{\qo}{\mathbf{q}_0}
\newcommand{\St}{\mathbf{\tilde{S}}}
\newcommand{\M}{\mathbf{M}}
\newcommand{\Mt}{\mathbf{\tilde{M}}}
\newcommand{\Mij}{M_{ij}}
\newcommand{\Mtij}{\tilde{M}_{ij}}

\newcommand{\up}{\begin{bmatrix} \ut\\ \Pp\end{bmatrix}}
\newcommand{\lmat}{\begin{bmatrix} -(\w^2+i\w\Gamma)\rho\I-\dfrac{\rho}{c^2}\g\g^{\mathbf{T}}+\g(\nabla\rho)^{\mathbf{T}} & &
\nabla-\dfrac{\g}{c^2}\\
-\nabla^{\mathbf{T}}-\dfrac{\g^{\mathbf{T}}}{c^2} & & -\dfrac{1}{\rho c^2}\end{bmatrix}}

\newcommand{\beq}{\begin{equation}}
\newcommand{\eeq}{\end{equation}}
\newcommand{\bea}{\begin{eqnarray}}
\newcommand{\eea}{\end{eqnarray}}

\slugcomment{To be submitted to {\it The Astrophysical Journal}}
\shortauthors{Schlottmann \& Kosovichev}
\shorttitle{Traveltime Kernels in Spherical Coordinates}

\begin{document}

\title{Helioseismic Fr\'{e}chet Traveltime Kernels in Spherical Coordinates}
\author{R. B. Schlottmann and A. G. Kosovichev}
\affil{W. W. Hansen Experimental Physics Laboratory, Stanford University}
\affil{Stanford, CA 94305}
\email{briansch@fusemail.com}

\begin{abstract}

Seismic traveltime measurements are a crucial tool in the investigation of the solar interior, particularly in the examination of
fine-scale structure.  Traditional analysis of traveltimes relies on a geometrical ray picture of acoustic wave propagation, which 
assumes high frequencies.  However, it is well-known that traveltimes obtained from finite-frequency waves are sensitive to variations 
of medium parameters in a wide Fresnel zone around the ray path.  To address this problem, Fr\'{e}chet traveltime sensitivity kernels 
have previously been developed.  These kernels use a more realistic approximation of the wave propagation to obtain a linear relationship between 
traveltimes and variations in medium parameters.  Fr\'{e}chet kernels take into account the actual frequency content of the measured 
waves and, thus, reproduce the Fresnel zone.  Kernel theory has been well-developed in previous work on plane-parallel models 
of the Sun for use in local helioseismology.  Our primary purpose is to apply kernel theory to much larger scales and in a spherical
geometry.  We also present kernel theory in a different way, using basic functional analytic methods, in the hope that this approach 
provides an even clearer understanding of the theory, as well as a set of tools for calculating kernels. Our results are very general
and can be used to develop kernels for sensitivity to sound speed, density, magnetic fields, fluid flows, and any other medium parameter
which can affect wave propagation.

\end{abstract}

\section{INTRODUCTION}

In both the Sun and the Earth, the passage of waves, acoustic or elastic, respectively, through the body yields an 
indispensable probe of the inner structure and physics that would otherwise be unavailable.  Both helioseismology and 
geoseismology have relied on the inversion of normal mode data and time-distance measurements to obtain information on
their subjects' interiors.  In both fields, the time-distance measurements are the data that are more sensitive to 
small-scale, three-dimensional variations in structural parameters such as wave speed and density.

Traditionally, the measurements of traveltimes have been interpreted using a geometrical ray approximation to the wave 
propagation.  However, it has long been appreciated that this approximation is generally inadequate, geometrical rays 
being accurate representations of waves only at very small wavelengths relative to variations in medium parameters.  In 
reality, the arrival time of a wave packet associated with a particular ray path is sensitive to structure well off the
ray, the width of that ``Fresnel zone'' being dependent on the dominant frequencies in the wave.

In an effort to obtain better information on the Earth's interior, geoseismologists developed traveltime sensitivity 
kernels \citep[See, e.g.,][]{DNH,HDN}, which account for the finite frequency nature of the waves.  By more accurately
representing the physics of wave propagation, these kernels provide a better ``map'' between structure and traveltimes,
which, in turn, yields a more accurate inversion result.

Traveltime kernels are not new to helioseismology.  \citet{GB} developed 2D kernels for f-mode traveltime sensitivity to
acoustic damping and excitation, while \citet{BKD}, using a plane-parallel model of the Sun's near-surface,  developed 3D 
traveltime kernels for perturbations in sound speed and density.  In both papers the theoretical foundations of sensitivity 
kernels, as well as practical considerations in their calculation, were laid out clearly and accurately.  We will rely most
heavily on those two sources to guide us in this work.

The primary purpose of this paper is to further generalize the work of \citet{GB} and \citet{BKD} to a spherical geometry, 
which would allow the development of kernels for deep Sun structure, an area of active interest \citep{Zhao} .  However, we also 
wish to present the theory using the tools of basic functional analysis, with the hope that this way of looking at kernels 
provides an even clearer understanding of their nature. 

Our theoretical results are quite general and can be used to develop kernels for various structural parameters.  Not only can
formulas for sensitivity to sound speed and density be obtained, but one can also derive kernels for magnetic fields, fluid
flows, and damping.  In principle, the sensitivity to any parameter which can affect the propagation of waves can be derived.

\section{INTRODUCTION TO KERNEL THEORY}\label{IntroKern}

In this section, we lay out a fairly general development of helioseismic traveltime sensitivity 
kernels based on the Born approximation.  In geoseismology, where their theory and application
were first developed, they are frequently called Fr\'{e}chet kernels or, more colorfully, ``banana-doughnut'' 
kernels, because of their characteristic appearance in cross-section.  As will be made precise later, these kernels quantify the 
first-order (i.e., linear) dependence of seismic traveltimes on perturbations of any physical properties of a medium 
that can affect the propagation of seismic waves.  By perturbation, we mean the difference between the value of a 
property for a particular medium and that of a reference medium.  Usually, the reference medium is taken to be
one with a high degree of symmetry, which greatly reduces the computational cost of calculating kernels.  In 
the case of both the Sun and the Earth, this means that the reference model is most often spherically symmetric.

The theoretical results contained in this section are not particularly novel.  Kernels have been developed previously for the Sun
in Cartesian coordinates and for the Earth (which has somewhat different wave propagation effects from the Sun)
in spherical coordinates.  However, the development presented here utilizes a different theoretical apparatus which may be 
clearer and cleaner in some ways than the usual derivations.  In particular, heavy use is made of some basic 
concepts in functional analysis.  These concepts are not complicated, but since they may not be familiar to every
reader, we give below a non-rigorous review of what will be needed.  \citep[See][ch. 9, for a nice review of functional
calculus.]{Hat}

\subsection{Basic Functional Analysis}\label{BFA}

First, we define a functional as a ``function over function space'' or, more precisely, a mapping from a space of functions
to a finite-dimensional space of real or complex numbers.  If $f$ is a functional of a function $g$ and 
itself a {\em function} of a variable $x$, we denote this as $f[g](x)$.  A relevant example would be the acoustic displacement
wavefield, $\ubf$, in a medium of sound speed $c$.  Since the field is a function of position and time as well as a functional
over the space of sound speed models, we would denote it here by $\ubf[c](\x,t)$.  

The workhorse in our use of functional analysis will be the functional derivative, which is formally (i.e., non-rigorously) 
defined as
\beq
\fder{f[g]}{g(a)}=\lim_{\ep\rightarrow 0}\frac{f[g(\cdot)+\ep\del{\cdot-a}]-f[g(\cdot)]}{\ep},
\eeq
where the dot in $g(\cdot)$, for example, is a placeholder for the unspecified argument of function $g$.
It may be useful to conceptualize the functional derivative as a gradient in an infinite dimensional space, with 
the value of the function at each point in physical space viewed as one coordinate of the function's ``position'' in 
function space. 

One consequence of our definition is that if 
\beq
f[g](x)=\fint dx^\prime g(x^\prime)\del{x-x^\prime})=g(x),
\eeq
then our definition of the functional derivative implies
\beq
\fder{f[g](x)}{g(a)}=\fder{g(x)}{g(a)}=\del{x-a},
\eeq
which is a result we will use often.


The only other concept from functional analysis we need is that of the functional Taylor series.  If $f$ is a 
functional over some function space, then $f$ can be expanded about some specific function $g_0$:
\beq
f[g]=f[g_0]+\fint dx^\prime \Delta g(x^\prime)\left.\fder{f[g]}{g(x^\prime)}\right|_{g_0}
+\frac{1}{2!}\fint dx^{\dprime} \fint dx^\prime \Delta g(x^\prime)\Delta g(x^{\dprime})
\left.\frac{\delta^2 f[g]}{\delta g(x^{\dprime})\delta g(x^\prime)}\right|_{g_0}+\cdots
\eeq
where
\beq
\Delta g(\cdot)=g(\cdot)-g_0(\cdot).
\eeq

Note that all the above concepts relationships extend in an obvious way to functionals over functions of 
several variables as well as to functionals over vector functions.

\subsection{Development of the Kernels}\label{DevKern}

Here we define the inverse problem to be solved and present the mathematical derivation of the
traveltime kernels that arise from it.  We first cover the definition of some basic observables, the 
precise definition of what we mean by ``traveltime'', and the linearization of the 
relationship between traveltimes and actual medium properties.  We then connect the results 
explicitly to wavefields in a reference solar model, providing general formulas for the kernels 
in terms of observational constraints (such as filters, line-of-sight effects, etc.), the reference wavefields, and 
the differential operators that govern them.

\subsubsection{Some Definitions and the Linearized Problem}\label{LinProb}

Some care is required here in our definition of various quantities of interest since there are different observables
which could be called the ``wavefield'' and different derived quantities which could be called the ``data''.  We 
will refer to the actual particle motion due to the propagation of helioseismic waves in the Sun as the ``wavefield''.  
The measured line-of-sight dopplergrams will be called the ``raw data'' or ``raw traces'', whereas the seismograms 
acquired by cross-correlating raw traces will be denoted ``processed traces'' or ``processed data''.

In addition, we make a couple of brief comments about notation.  To promote clarity, we indicate solar surface
coordinates and integration variables by $\s$, $\sgp$, and so forth, whereas interior coordinates and integration
variables are denoted by $\x$ and its variations.  Also, unless otherwise stated, the Einstein summation convention is
implied by repeated Latin indices.

We denote the particle velocity wavefield by $\vb(\x,t)$ and the raw data by $\psi(\s,t)$.  These two are related by
\beq
\psi(\s,t)=\ssi \!\! d^2\sgp \!\!\fint \!\!\! d\tp \,\F(\s,\sgp,t-\tp)\cdot\vb(\sgp,\tp)\label{Psioft}
\eeq
where $\F$ represents the action of all spatial and temporal filters, which are assumed to be linear and time-translation
invariant, that result from the instrument response and from whatever other filters that may be applied later.  Note that the 
direction of $\F$ is along the line of sight from the point $\sgp$ to the instrument and that the integration is over the 
surface of the Sun.

The processed traces derived from the raw traces are denoted $D(\sga,\sgb,t)$ and are obtained by cross-correlation of the raw
traces associated with points $\sga$ and $\sgb$ on the solar surface:
\beq
D(\sga,\sgb,t)=\fint \!\!\! d\tp \psi(\sga,t+\tp)\psi(\sgb,\tp).\label{Doft}
\eeq  
For simplicity, we will not account for the finite time length of raw traces  but instead will 
approximate the cross-correlation as an integral over all time.  Note that we have defined 
$D(\sga,\sgb,t)$ in such a way that for positive times, the resulting seismograms represent waves travelling from $\sgb$ to $\sga$.

We also define the quantities $\vb_0$, $\psi_0$, and $D_0$ as the wavefield, raw data, and processed data, respectively,
generated synthetically from a given reference solar model.  Hereafter, any quantity with a subscript $0$ is taken to be
associated with the reference model.

We next define $C(\sga,\sgb,t)$ as the windowed cross-correlation between the real and reference processed traces:
\beq
C(\sga,\sgb,t)=\fint \!\!\! d\tp W(\sga,\sgb,t+\tp)D_0(\sga,\sgb,t+\tp)W(\sga,\sgb,\tp)D(\sga,\sgb,\tp).\label{Coft}
\eeq
The function $W$ is a windowing function applied to both the real and reference data, the purpose of which is to isolate
pulses associated with specific propagation paths.

Now let $\tau(\sga,\sgb)$ be the time lag at which the cross-correlation above is maximum.  We define this time as the 
traveltime.  It is imperative to note that the traveltime {\em must} be measured this way before the kernels we derive 
here can be used.  Other ways of measuring the traveltime are not necessarily better or worse, but they will require different 
kernels.  We also point out that the way we have defined the cross-correlation ensures that if a pulse in the real data is early 
with respect to the reference data, then the traveltime will be {\em positive}.  In other words, $\tau$ is a measure of how much 
the real pulse is advanced with respect to the reference.

The traveltime is a functional of the actual solar model.  If we let $\q$ and $\qo$ denote vectors of functions describing
the actual and reference solar models (e.g. $q_1(\x)$ and $q_2(\x)$ could be, respectively, the sound speed and density as a function 
of space inside the real Sun), then we can expand $\tau$ in a functional Taylor series about $\qo$,
\begin{align}
\tau[\q](\sga,\sgb)&=\tau[\qo](\sga,\sgb)+\sum_\alpha\svi\!\!d^3\x\ \Delta q_\alpha(\x)
\left.\fder{\tau}{q_\alpha(\x)}(\sga,\sgb)\right|_{\qo}\nonumber\\
&\hspace{20mm}+\frac{1}{2!}\sum_{\alpha,\beta}\svi\!\!d^3\x\!\!\svi\!\!d^3\xp\Delta q_\alpha(\x)\Delta q_\beta(\xp)
\left.\frac{\delta^2\tau}{\partial q_\beta(\xp)\partial q_\alpha(\x)}(\sga,\sgb)\right|_{\qo}+\cdots,
\end{align}
where $\Delta q_\alpha(\x)=q_\alpha(\x)-q_{0,\alpha}(\x)$, and integration is over the volume of the Sun.  The first term in the 
expansion is identically zero since the time lag between the real and reference data would be zero if the real model were the same as 
the reference.  In order to get a linear inverse problem to be solved later, we need a linear relationship between the traveltime
and the perturbations $\Delta q_\alpha$ to the solar model.  This is accomplished by simply dropping the nonlinear terms in the 
above Taylor series.  This is the {\em only} linearization required in this development of the kernels.  Any other linearizations
that occur later (e.g. the use of the first Born approximation) are a direct consequence of this one approximation and do
not represent further approximations.  However, this linearization does assume that the model perturbations are small in the
sense that their main effect is to cause small changes in the arrival times of pulses in the real data.  Larger perturbations
invalidate this assumption because they can cause distortions in waveforms, not just time shifts, that make the cross-correlation
traveltime measurements meaningless.  Thus, hereafter we assume the equivalence
\beq
\tau(\sga,\sgb)\equiv\sum_\alpha\svi\!\!d^3\x\ \Delta q_\alpha(\x)\left.\fder{\tau}{q_\alpha(\x)}(\sga,\sgb)\right|_{\qo}.\label{LinApprox}
\eeq

Traditionally, Fr\'{e}chet kernels are defined in such a way that they provide a linear relationship between traveltimes
and {\em fractional} perturbations in model parameters.  Following this tradition, we define $K_\alpha$, the kernel for the $\alpha$-th
model parameter, as 
\beq
K_\alpha(\x;\sga,\sgb)=q_{0,\alpha}(\x)\left.\fder{\tau}{q_\alpha(\x)}(\sga,\sgb)\right|_{\qo},\label{KerDef}
\eeq
so that
\beq
\tau(\sga,\sgb)\equiv\sum_\alpha\svi\!\!d^3\x\ \frac{\Delta q_\alpha(\x)}{q_{0,\alpha}(\x)}K_\alpha(\x;\sga,\sgb).
\eeq
The remainder of this section will be devoted to relating the above expression, eq.~(\ref{KerDef}), to calculable physical 
quantities and known observational constraints. 

\subsubsection{General Kernel Formulas}\label{GenKern}

The expression in eq.~(\ref{KerDef}) is useless until we connect it explicitly to the underlying physics governing helioseismic wave 
propagation.  We begin by providing an explicit mathematical definition of $\tau$.  We have previously defined it as the value of the 
time lag that maximizes $C$, the windowed cross-correlation between the real and reference data.  Letting $C(t)=C(\sga,\sgb,t)$, we
expand $C$ in a Taylor series about $t=0$:
\beq
C(t)=\sum_{n=0}^{\infty}\frac{t^n}{n!}C^{(n)}(0).
\eeq
Making the reasonable assumption that $C$ is a smooth function, we have that $C^{\prime}(\tau)=0$.  Thus,
\beq
C^{\prime}(\tau)=\sum_{n=1}^{\infty}\frac{\tau^{n-1}}{(n-1)!}C^{(n)}(0)=0.\label{Cp}
\eeq 
Taking the first functional derivative of eq.~(\ref{Cp}) with respect to $q_\alpha$ at $\q=\qo$, we get
\beq
0=\left.\fder{C^{\prime}(\tau)}{q_\alpha(\x)}\right|_{\qo}\!\!\!\!=\ \ \left.\fder{C^{\prime}(0)}{q_\alpha(\x)}\right|_{\qo}\!\!\!\!
+\ \ \sum_{n=2}^{\infty}\left.\left[\frac{\tau^{n-2}}{(n-2)!}C^{(n)}(0)\fder{\tau}{q_\alpha(\x)}
+\frac{\tau^{n-1}}{(n-1)!}\fder{C^{(n)}(0)}{q_\alpha(\x)}\right]\right|_{\qo}.
\eeq
Since $\tau[\qo]=0$, the above reduces to
\beq
0=\left.\fder{C^{\prime}(0)}{q_\alpha(\x)}\right|_{\qo}\!\!\!\!+\ \ C_0^{\dprime}(0)\left.\fder{\tau}{q_\alpha(\x)}\right|_{\qo},
\eeq
which implies
\beq
\left.\fder{\tau}{q_\alpha(\x)}\right|_{\qo}=-\frac{1}{C_0^{\dprime}(0)}
\left.\fder{C^{\prime}(0)}{q_\alpha(\x)}\right|_{\qo},\label{dtaudq}
\eeq
where 
\beq
C_0(t)=\fint \!\!\! d\tp W(\sga,\sgb,t+\tp)D_0(\sga,\sgb,t+\tp)W(\sga,\sgb,\tp)D_0(\sga,\sgb,\tp).
\eeq
Eq.~(\ref{dtaudq}) connects our definition of the kernel in eq.~(\ref{KerDef}) to something readily related to actual wave
propagation.

At this point, it will be more convenient to work in the frequency domain.  To begin, we state our Fourier transform 
convention.  If $f(t)$ is a function of time, then
\begin{align}
\addtocounter{equation}{1}
\tilde{f}(\w)&=\frac{1}{\sqrt{2\pi}}\fint dt e^{i\w t}f(t)\label{FTa}\tag{\theequation a}\\
f(t)&=\frac{1}{\sqrt{2\pi}}\fint d\w e^{-i\w t}\tilde{f}(\w).\label{FTb}\tag{\theequation b}
\end{align}
Now, if we define
\beq
U_n(\sga,\sgb,\tp)\equiv\frac{d^n}{dt^n}\Big[W(\sga,\sgb,t+\tp)D_0(\sga,\sgb,t+\tp)\Big]_{t=0}W(\sga,\sgb,\tp),\label{U12}
\eeq
where $n=1,2$, then from eq.~(\ref{Coft})
\begin{align}
\addtocounter{equation}{1}
C^\prime(0)&=\fint d\tp\ U_1(\sga,\sgb,\tp)D(\sga,\sgb,\tp)\tag{\theequation a}\\
&=\fint d\w\ \tilde{U}_1^{*}(\sga,\sgb,\w)\tilde{D}(\sga,\sgb,\w),\tag{\theequation b}\label{Cp0}
\end{align}
and, similarly,
\beq
C_0^\dprime(0)=\fint d\w\ \tilde{U}_2^{*}(\sga,\sgb,\w)\tilde{D}_0(\sga,\sgb,\w),\label{Cpp}
\eeq
where we have used the fact that $U_{1,2}$ and $D$ are real-valued.  With eq.~(\ref{Cp0}) in hand, we find
\beq
\left.\fder{C^{\prime}(0)}{q_\alpha(\x)}\right|_{\qo}\!\!\!\!=\ \,\fint d\w\ \tilde{U}_1^{*}(\sga,\sgb,\w)
\left.\fder{\tilde{D}(\sga,\sgb,\w)}{q_\alpha(\x)}\right|_{\qo}.\label{dCpdq}
\eeq
Thus, combining eqs.~(\ref{KerDef}), (\ref{dtaudq}), and (\ref{dCpdq}), we have as an intermediate result
\beq
K_\alpha(\x;\sga,\sgb)=-\frac{q_{0,\alpha}(\x)}{C_0^{\dprime}(0)}\ \,\fint d\w\ \tilde{U}_1^{*}(\sga,\sgb,\w)
\left.\fder{\tilde{D}(\sga,\sgb,\w)}{q_\alpha(\x)}\right|_{\qo}.\label{KDressed}
\eeq

Since we will deal only with linearized wave propagation, we know the wavefield obeys
\beq
\Ls\Up(\x,\w)=\St(\x,\w)\label{LUpS}
\eeq
where $\Ls$ is some linear differential operator that depends on the model parameters $\q$, $\St$ is a source function, 
and $\Up$ is a four-dimensional vector comprised of the particle displacement wavefield $\ut$ and the pressure perturbation $\Pp$.
If $\Gt(\x,\w;\xo)$ is the Green's tensor in the actual solar model, i.e., if 
\beq
\Ls\Gt(\x,\w;\xo)=\del{\x-\xo}\I,\label{Geq}
\eeq
where $\I$ is the identity matrix, then the solution to eq.~(\ref{LUpS}) can be written as
\beq
\Up(\x,\w)=\svi d^3\xp \Gt(\x,\w;\xp)\cdot\St(\xp,\w).
\eeq
With this, eq.~(\ref{Psioft}) can be rewritten in the frequency domain as
\beq
\tilde{\psi}(\s,\w)=\sqrt{2\pi}\ssi d^2\sgp \svi d^3\xp \tilde{F}_i(\s,\s^\prime,\w)\Gtij(\s^\prime,\w;\xp)\tilde{S}_j(\xp,\w),
\eeq
where we take $\Ft$ to have zero pressure component.  Using this in the definition of $D$, eq.~(\ref{Doft}), we get
\begin{align}
\tilde{D}(\sga,\sgb,\w)&=\sqrt{2\pi}\,\tilde{\psi}(\sga,\w)\,\tilde{\psi}^{*}(\sgb,\w)\notag\\
&=\ssi d^2\s_1\svi d^3\x_1 \ssi d^2\s_2\svi d^3\x_2\, \tilde{F}_i(\sga,\s_1,\w)\Gtij(\s_1,\w;\x_1)\tilde{S}_j(\x_1,\w)\notag\\ 
&\hspace{60mm}\times\tilde{F}^{*}_k(\sgb,\s_2,\w)\Gtkl^{*}(\s_2,\w;\x_2)\tilde{S}^{*}_l(\x_2,\w).\label{DSS}
\end{align}

Regarding the source terms appearing above, we note that it is understood that the acoustic excitation in the Sun is due to 
turbulent convection just below the photosphere and are often modeled stochastically.  Following \citet{BKD}, we
choose to represent the excitation by a source covariance matrix, $\M$, defined as
\beq 
\Mij(\x,\xp,t)=\fint d\tp S_i(\x,t+\tp)S_j(\xp,\tp)\label{M},
\eeq
or, in the frequency domain,
\beq
\Mtij(\x,\xp,\w)=\tilde{S}_i(\x,\w)\tilde{S}^{*}_j(\xp,\w).\label{Mt}
\eeq
(The value of using this representation will come later when we use a fairly simple reference model of the ensemble average of $\Mt$. 
However, we will also consider the possibility that the true form of $\Mt$ could be taken as one of the 
model parameters in $\q$.)  Using this definition, eq.~(\ref{DSS}) becomes
\begin{align}
\tilde{D}(\sga,\sgb,\w)&=\ssi d^2\s_1\svi d^3\x_1 \ssi d^2\s_2\svi d^3\x_2\, \tilde{F}_i(\sga,\s_1,\w)\Gtij(\s_1,\w;\x_1)\notag\\
&\hspace{60mm}\times\tilde{F}^{*}_k(\sgb,\s_2,\w)\Gtkl^{*}(\s_2,\w;\x_2)\tilde{M}_{jl}(\x_1,\x_2,\w)\label{DMt}
\end{align}

Within eq.~(\ref{DMt}), only the Green's functions and the source covariance matrix are ever dependent on the model parameters.
Thus, for simplicity, when we take the functional derivative of $\tilde{D}$, we separate those out and define the quantity
\beq
\tilde{R}_{\alpha,ik}(\x;\s_1,\s_2,\w)=\left.\fder{}{q_\alpha(\x)}\svi d^3\x_1 \svi d^3\x_2\,
\Gtij(\s_1,\w;\x_1)\Gtkl^{*}(\s_2,\w;\x_2)\tilde{M}_{jl}(\x_1,\x_2,\w)\right|_{\qo}\label{RawKer}.
\eeq
In order to evaluate this, we will need the functional derivative of $\G$.  We start by taking the 
derivative of both sides of eq.~(\ref{Geq}), bearing in mind that the right hand side does not depend on the model parameters,
\beq
\left.\fder{\Ls}{q_\alpha(\x)}\right|_{\qo}\!\!\!\!\!\Gto(\xp,\w;\xo)
+\left.\Lo\fder{\Gt(\xp,\w;\xo)}{q_\alpha(\x)}\right|_{\qo}\!\!\!\!\!=\ \ 0,
\eeq
where $\Lo$ and $\Gto$ are the wave equation differential operator and Green's function, respectively, for the reference model.
The solution to this equation is 
\beq
\left.\fder{\Gt(\xp,\w;\xo)}{q_\alpha(\x)}\right|_{\qo}=-\svi d^3\xpp \Gto(\xp,\w;\xpp)
\cdot\left[\left.\fder{\Ls(\xpp,\w)}{q_\alpha(\x)}\right|_{\qo}\!\!\!\!\!\Gto(\xpp,\w;\xo)\right].
\eeq

Putting all of our results together so far, we quote our general equations for the desired 
Fr\'{e}chet traveltime kernels:
\begin{eqnarray}
K_{\alpha}(\x;\sga,\sgb)\!\!\!&=\!\!\!&\frac{-q_{0,\alpha}(\x)}{C_0^\dprime(\sga,\sgb,0)}
\fint\!\!\! d\w \ssi\!\!\! d^2\s_1 \ssi\!\!\! d^2\s_2
\tilde{U}_1^{*}(\sga,\sgb,\w)\tilde{F}_i(\sga,\s_1,\w)\tilde{F}^{*}_k(\sgb,\s_2,\w)\nonumber\\
& &\hspace{60mm}\times\tilde{R}_{\alpha,ik}(\x;\s_1,\s_2,\w)\label{KerFilt}
\end{eqnarray}
where
\begin{eqnarray}
\tilde{R}_{\alpha,ik}(\x;\s_1,\s_2,\w)\!\!\!\!\!&=\!\!\!\!\!&\svi d^3\x_1 \svi d^3\x_2
\left\{\tilde{G}_{0,ij}(\s_1,\w;\x_1)\tilde{G}_{0,kl}^{*}(\s_2,\w;\x_2)
\left.\fder{\tilde{M}_{jl}(\x_1,\x_2,\w)}{q_{\alpha}(\x)}\right|_{\qo}\right.\nonumber\\
& &\hspace{-15mm}\left.-\svi\!\! d^3\xpp\!\!\left[
\!\tilde{G}_{0,im}(\s_1,\w;\xpp)
\!\!\left(\!\!\left.\fder{\Ls(\xpp,\w)}{q_\alpha(\x)}\right|_{\qo}\!\!\Gto(\xpp,\w;\x_1)\!\!\right)_{mj}
\!\!\tilde{G}_{0,kl}^{*}(\s_2,\w;\x_2)\right.\right.\nonumber\\
& &\hspace{-30mm}\left.\left.+\tilde{G}_{0,ij}(\s_1,\w;\x_1)\tilde{G}^{*}_{0,kn}(\s_2,\w;\xpp)
\!\!\left(\!\!\left.\fder{\Ls(\xpp,\w)}{q_\alpha(\x)}\right|_{\qo}\!\!\Gto(\xpp,\w;\x_2)\!\!\right)^{*}_{nl}
\right]\tilde{M}_{0,jl}(\x_1,\x_2,\w)\right\}.\label{FullRawKer}
\end{eqnarray}

We note that the presence of three coupled volumetric integrals in eq.~(\ref{FullRawKer}) would seem to make the calculation
of the raw kernel computationally impractical.  However, the functional derivatives of the linear wave operator generally will be
proportional to three-dimensional spatial delta functions and their derivatives.  Additionally, the reference source covariance 
matrix can be taken to be proportional to delta functions.  Thus, the actual computational burden is much smaller
than one might fear from the above result.  We will see this in specific examples for kernels, such as those
for sound speed and density. 

\section{SOME SPECIFIC KERNEL FORMULAS}\label{SpecK}

In this section, we specify the aspects of acoustic wave propagation that we will consider and apply our general kernel formulas to 
a particular reference model of the Sun's structure and ensemble average of the source covariance matrix.  We then 
give formulas for sensitivity kernels for perturbations in sound speed squared and density.

\subsection{The Wave Equation and Reference Model}

In order to use the kernel formulas we have developed, we must make assumptions about the physics of acoustic wave propagation 
both in the Sun and in our reference model.  Only then can we calculate our reference Green's functions and the functional 
derivatives of the wave equation operator that are needed.

From here on, we will assume that there are no magnetic fields or bulk fluid flows present in the Sun.  Additionally, we will work
in the Cowling approximation, i.e., we assume that the density variations induced by the propagation of an acoustic wave have a 
negligible effect on the gravitational potential.  For all but the lowest degree modes, this approximation is extremely accurate.
In future work, we will relax most if not all of these restrictions.

We will also include damping in our wave equation.  For simplicity, we will somewhat follow \citet{BKD} and use a 
simple convolutional model in the time domain.  However, in contrast to their work, we will assume that the attenuation operator is
spatially local.

As for boundary conditions, we again follow \citet{BKD} and use a free surface boundary condition, i.e., we set the Lagrangian pressure equal to
zero at some radius.  As noted in \citet{BKD}, this does, of course, mean that waves with frequencies above the acoustic cutoff do not
escape the Sun but are reflected back into it, in disagreement with reality.

Under the above assumptions, the linearized equations for particle displacement $\ut$ and pressure perturbation $\Pp$ have the
following well-known form in the frequency domain: 
\begin{gather}
-\w^2\rho\ut+\nabla\Pp-\g\left[\frac{\Pp}{c^2}+\frac{\rho}{c^2}\g\cdot\ut-(\nabla\rho)\cdot\ut\right]-i\w\rho\Gamma\ut=0,\label{NewtII}\\
\Pp+c^2\rho\dvg\ut+\rho\g\cdot\ut=0,\label{PpOrig}
\end{gather}
where $c$ and $\rho$ are the sound speed and density, respectively, $\g$ is the gravitational acceleration, $\omega$ is the angular 
frequency, and $\Gamma$ is the damping parameter.  For the real Sun, we assume that all of the medium parameters above can vary 
spatially in all three dimensions but not in time.  In the case of the damping parameter, we assume that it also varies in frequency 
so that in the time domain it is convolved with the particle velocity $\vb$.

However, for the purpose of constructing our wave equation operator $\Ls$, we will rewrite eq.~(\ref{PpOrig}) slightly:
\beq
-\frac{1}{\rho c^2}\Pp-\dvg\ut-\frac{\g}{c^2}\cdot\ut=0\label{PpFinal}.
\eeq 
With this modification, our wave equation, eq.~(\ref{LUpS}), becomes
\beq
\Ls\up=\lmat\up=\St,
\eeq
where all vector quantities (and $\nabla$) are taken to be column vectors and $\mathbf{T}$ indicates transposition.  The reason for 
using eq.~(\ref{PpFinal}) is that in this form, the Green's matrix corresponding to $\Ls$ satisfies reciprocity, i.e., 
\beq
\Gt(\x,\w;\xo)=\Gt^{\mathbf{T}}(\xo,\w;\x),
\eeq
Having a Green's matrix that obeys reciprocity greatly reduces the computational cost 
of computing kernels by limiting the number of source points for which we need to calculate responses, at least when one uses a technique
other than mode summation, as we do.

With a specific form of $\Ls$ now given, we are in position to calculate its functional derivatives.  In this paper, we will concern
ourselves with kernels for sound speed squared and density, so we only need the functional derivatives with respect to those quantities:
\begin{gather}
\left.\fder{\Ls(\x)}{c^2(\xp)}\right|_{\qo}=\frac{1}{c^4_0}\del{\x-\xp}\begin{bmatrix}\rho_0\go\go^{\mathbf{T}} &  & \go \\
\go^{\mathbf{T}} & & \dfrac{1}{\rho_0}\end{bmatrix},\label{dLdc}\\
\mbox{}\nonumber\\
\left.\fder{\Ls(\x)}{\rho(\xp)}\right|_{\qo}=\del{\x-\xp}\begin{bmatrix}-(\w^2+i\w\Gamma_0)\I-\dfrac{1}{c^2_0}\go\go^{\mathbf{T}} & & 0\\
0 & & \dfrac{1}{\rho^2_0c^2_0}\end{bmatrix}+\begin{bmatrix} \go\nabla^{\mathbf{T}}\del{\x-\xp} & & 0\\ 0 & & 0\end{bmatrix}.\label{dLdrho}
\end{gather}
In eq.~(\ref{dLdrho}) we have neglected the derivative of $\g$ with respect to density under the assumption that the deviations in
density from our reference model will cause such small changes in gravity that they will have a negligible effect on wave propagation.
Note that this assumption is separate from the Cowling approximation.  

In addition to laying out our assumptions about wave propagation in the Sun, we must also describe our reference model.  The obvious choice
for a model of the Sun's structure is one that is spherically symmetric and in hydrostatic equilibrium. The symmetry of this choice greatly 
simplifies the calculation of the Green's functions we will need.  

As for the reference model of acoustic excitation, we will assume a simple form for the ensemble average of the source covariance matrix, 
eq.~(\ref{Mt}).  We choose to generalize \citet{BKD} model to spherical coordinates:
\beq
\left<\Mtij(\x,\xp,\w)\right>=\delta_{ir}\delta_{jr}\frac{\del{\theta-\theta^\prime}\del{\phi-\phi^\prime}}{r^2r^{\prime 2}\sin\theta}
\delta^\prime\left(r-r_s\right)\delta^\prime\left(r^\prime-r_s\right)\left|\tilde{f}(\w)\right|^2,\label{SC}
\eeq
where a subscript $r$ indicates a component in the radial direction, and $|\tilde{f}(\w)|^2$ is the power spectrum of 
the excitation and may be adjusted to ensure that the waves modelled in 
the reference medium have an acoustic power spectrum close to the Sun's.  Note that this a very practical model of excitation, not 
necessarily a realistic one.  In particular, the presence of the $\delta$-functions in the angular variables, which implies a zero 
horizontal correlation length for the sources, simplifies the computational effort required to model the excitation.  As noted 
in \citet{GB}, the true correlation length of the sources is much smaller than the wavelengths of the waves we will 
model, anyway. 

\subsection{Kernel Formulas for Perturbations to $c^2$ and $\rho$}

With our above assumptions in hand, we can now construct our sound speed and density kernel formulas.  To begin, we introduce some 
simplifying notation.  Recalling that our Green's function is a $4\times4$ matrix corresponding to a coupled system of equations in 
both particle displacement and pressure perturbation, we indicate its components by the indices $r$, $\theta$, $\phi$, and $p$, denoting 
the three directions in spherical coordinates and pressure.  For instance, $G_{0,rp}(\x,\w;\xo)$ denotes the radial component of 
displacement at point $\x$ due to a point pressure source at $\xo$, in the reference medium, at frequency $\w$.  When we wish to leave
a component unspecified (or indicate Einstein summation), we use Latin indices (other than $r$ or $p$).

To construct our kernel formulas, we start by incorporating the source excitation model, eq.~(\ref{SC}), into eq.~(\ref{FullRawKer}),
which we write as
\beq
\tilde{R}_{\alpha,ik}(\x;\s_1,\s_2,\w)=H_{\alpha,ik}(\x;\s_1,\s_2,\w)+H^{*}_{\alpha,ki}(\x;\s_2,\s_1,\w)\label{RHH}
\eeq
where
\begin{align}
H_{\alpha,ik}(\x;\s_1,\s_2,\w)&=\nonumber\\
&\hspace{-27mm}-\!\!\ssi\!\!\!d^2\!\sgp\!\!\!\svi\!\!\! d^3\xpp\tilde{G}_{0,im}(\s_1,\w;\xpp)\!\!
\left(\left.\fder{\Ls(\xpp,\w)}{q_\alpha(\x)}\right|_{\qo}\!\!\partial_{r_s}\Gto(\xpp,\w;r_s,\sgp)\right)_{mr}\!\!\!
\partial_{r_s}\tilde{G}_{0,kr}^{*}(\s_2,\w;r_s,\sgp)\label{Halph}
\end{align}
Note that $\partial_{r_s}$ is shorthand for partial differentiation with respect to $r_s$, and we have assumed that the source covariance
is fixed.

Now using eqs.~(\ref{dLdc}) and (\ref{dLdrho}) and recognizing that $\go=-g_0\hat{r}$, we get
\begin{align}
H_{c^2}(\x;\s_1,\s_2,\w)&=\nonumber\\
&\hspace{-19mm}-\frac{1}{c^4_0}\!\!\ssi\!\!\!d^2\!\sgp\Biggl[\tilde{G}_{0,rr}(\s_1,\w;\x)
\left(\rho_0g^2_0\partial_{r_s}\tilde{G}_{0,rr}(\x,\w;r_s,\sgp)
-g_0\partial_{r_s}\tilde{G}_{0,pr}(\x,\w;r_s,\sgp)\right)\Biggr.\nonumber\\
&\hspace{0mm}+\tilde{G}_{0,rp}(\s_1,\w;\x)
\Biggl.\left(-g_0\partial_{r_s}\tilde{G}_{0,rr}(\x,\w;r_s,\sgp)
+\frac{1}{\rho_0}\partial_{r_s}\tilde{G}_{0,pr}(\x,\w;r_s,\sgp)\right)\Biggr]\nonumber\\
&\hspace{30mm}\times\partial_{r_s}\tilde{G}_{0,rr}^{*}(\s_2,\w;r_s,\sgp)\left|\tilde{f}(\w)\right|^2\label{Hc2}
\end{align}
and 
\begin{align}
H_{\rho}(\x;\s_1,\s_2,\w)&=-\ssi\!\!\!d^2\!\sgp\Biggl[
-(\w^2+i\w\Gamma)\sum_{m=r,\theta,\phi}\tilde{G}_{0,rm}(\s_1,\w;\x)\partial_{r_s}\tilde{G}_{0,mr}(\x,\w;r_s,\sgp)\nonumber\\
&\hspace{-12mm}-\tilde{G}_{0,rr}(\s_1,\w;\x)\frac{g^2_0}{c^2_0}\partial_{r_s}\tilde{G}_{0,rr}(\x,\w;r_s,\sgp)
+\tilde{G}_{0,rp}(\s_1,\w;\x)\frac{1}{\rho^2_0c^2_0}\partial_{r_s}\tilde{G}_{0,pr}(\x,\w;r_s,\sgp)\nonumber\\
&\hspace{0mm}+\partial_r\left(\tilde{G}_{0,rr}(\s_1,\w;\x)g_0\partial_{r_s}\tilde{G}_{0,rr}(\x,\w;r_s,\sgp)\right)\nonumber\\
&+\frac{1}{r}\partial_{\theta}\left(\tilde{G}_{0,rr}(\s_1,\w;\x)g_0\partial_{r_s}\tilde{G}_{0,\theta r}(\x,\w;r_s,\sgp)\right)\nonumber\\
&+\frac{1}{r\sin\theta}\partial_{\phi}\left(\tilde{G}_{0,rr}(\s_1,\w;\x)g_0\partial_{r_s}\tilde{G}_{0,\phi r}(\x,\w;r_s,\sgp)
\right)\Biggr]\nonumber\\
&\times\partial_{r_s}\tilde{G}_{0,rr}^{*}(\s_2,\w;r_s,\sgp)\left|\tilde{f}(\w)\right|^2,\label{Hrho}
\end{align}
where all medium parameters are implicitly evaluated at $\x$.

Lastly, we need to incorporate our source covariance model into our formula for $\tilde{D}_0$, the reference model version
of eq.~(\ref{DSS}), which is needed in the evaluation of $U_{1,2}$ and $C^\dprime_0$ (eq.~\ref{Cpp}), which in turn are used in
eq.~(\ref{KerFilt}).  Using eq.~(\ref{SC}), our formula for $\tilde{D}_0$ becomes
\begin{align}
\tilde{D_0}(\sga,\sgb,\w)&=\ssi d^2\s_1 \ssi d^2\s_2\ssi d^2\sgp, 
\tilde{F}_i(\sga,\s_1,\w)\partial_{r_s}\tilde{G}_{0,ir}(\s_1,\w;r_s,\sgp)\notag\\ 
&\hspace{40mm}\times\tilde{F}^{*}_k(\sgb,\s_2,\w)\partial_{r_s}\tilde{G}^{*}_{0,kr}(\s_2,\w;r_s,\sgp)\left|\tilde{f}(\w)\right|^2.\label{D0}
\end{align}

Thus, to construct traveltime sensitivity kernels for squared sound speed and density, under the assumptions laid out above, the equations 
that are needed are (\ref{U12}), (\ref{Cpp}), (\ref{KerFilt}), (\ref{RHH}), and (\ref{Hc2})-(\ref{D0}).  Of course, the user of these 
results will still need to provide additional information, such as a specific reference model, observational geometry, and so forth.

\section{EXAMPLES}\label{Examples}

In this section we display examples of kernels for squared sound speed for large distances, in order to give an idea of the extent
and character of traveltime sensitivity at depth in the Sun.  In order to accelerate the calculations somewhat, we make one particularly 
simplifying assumption, i.e.,  we take our filter function $\F$ (see eq.~\ref{Psioft}) to be  
\beq
\F(\s,\sgp,t)=\hat{r}\del{\s-\sgp}\del{t}\label{FSimple}.
\eeq
The assumption that the line of sight is always radial is quite helpful in lessening the computational complexity, but it is,
of course, only a good approximation near the center of the Sun's visible disk. 

As for our reference model, we use model S \citep{CD} as the model of the Sun's structure, and a free-surface boundary condition, as 
mentioned earlier.  For the power spectrum of the acoustic excitation, we once again turn to \citet{BKD} and use
\beq
\left|\tilde{f}(\w)\right|^2=e^{-w^2T^2_{src}}
\eeq
with $T_{src}=68$ s.

Our damping model is actually a fusion of the ones used in \citet{GB} and \citet{BKD}.  We let
\beq
\Gamma(\w,r)=g(\w)h(r),
\eeq
where from \citet{GB} we take
\beq
g(\w)=\gamma\left|\frac{\w}{\w_*}\right|^\beta
\eeq
with $\gamma/2\pi=100\mu$Hz, $\w_*/2\pi=3$mHz, and $\beta=4.4$.  Whereas from \citet{BKD} we get
\beq
h(r)=\exp\left[-\frac{(T(r)-T_c)^2}{(\Delta t)^2}\right]
\eeq
with 
\beq
T(r)=\int_0^r \frac{dr^\prime}{c_0(r^\prime)},
\eeq
$T_c=T(R)$, where $R$ is the radius of the photosphere, and $\Delta t=$69 s.

Lastly, we assume that our time window $W$ is simply one for times within 10 m of a first arrival and zero otherwise.

In Fig.~1, we see our first example.  Shown is a kernel for squared sound speed for two observation points separated by 
and angle of $30^{\mathrm{o}}$.  Because of the huge range of amplitudes in the kernel, the strongest regions being close
to the observation points, we have scaled the kernel by the sound speed and saturated the scale in order to bring out the
deep structure.  Indicated in both cross-sections is the location of the ray path associated with the first arrival at this 
distance.  Note that the kernel displays the classic ``banana-doughnut'' shape, with sensitivity reaching a local minimum
on the ray path.  The images here are limited to the upper 40\% of the Sun, by radius.  We can see that the kernel has 
sensitivity mostly in the convection zone and above.

In Fig.~2, we have the squared sound speed kernel for a distance of $60^{\mathrm{o}}$.  In this case, the sections show the upper
70\% of the Sun.  Clearly, the kernel in this case significantly intersects the tachocline, indicating that it may be of some use in
interpreting time-distance measurements aimed at studying that region.  

In both figures, it is clear that ray theory cannot be a good approximation here, given that the waves are clearly sensitive to 
a broad volume around the ray.  

\begin{figure}
\figurenum{1}
\epsscale{0.85}
\plotone{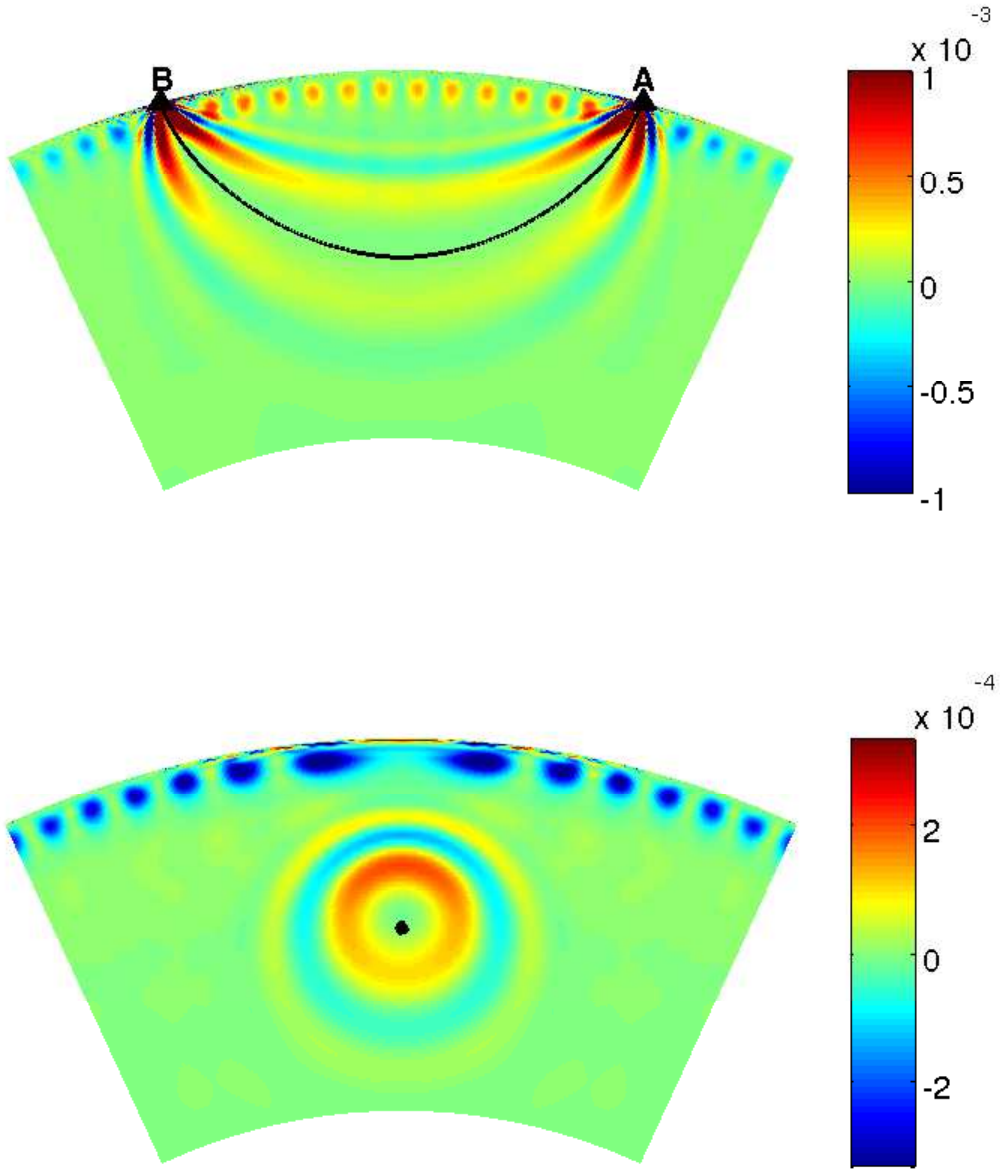}
\caption{Vertical cross-sections of a sensitivity kernel for squared sound speed for a distance of $30^{\mathrm{o}}$. The kernel has been
scaled by the sound speed and the scale has been severely saturated. (\emph{Top}) The 
cross-section in the plane of the ray path (black line).  (\emph{Bottom}) The cross-section perpendicular to the ray path.  The intersection
point of the plane and the ray is indicated by the black dot.  The total angular range in both sections is $50^{\mathrm{o}}$, and both
extend radially from 0.6 $R_\sun$ to the surface. Note that this is the kernel for waves travelling from point B to point A. The units are
Mm$^{-2}$} 
\end{figure}

\begin{figure}
\figurenum{2}
\epsscale{0.85}
\plotone{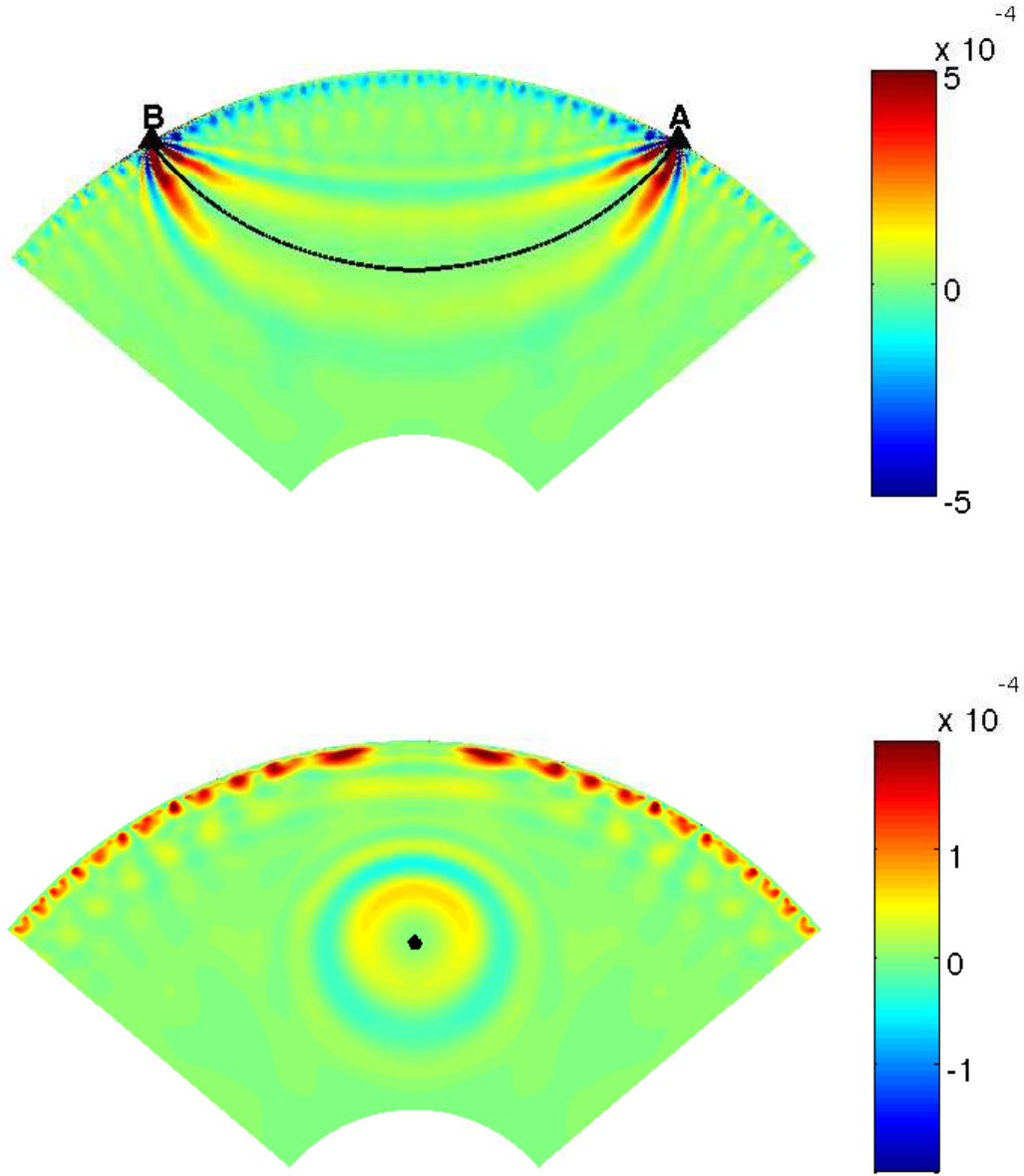}
\caption{Vertical cross-sections of a sensitivity kernel for squared sound speed for a distance of $60^{\mathrm{o}}$. Again the kernel has been
scaled by the sound speed and the scale has been saturated. (\emph{Top}) The 
cross-section in the plane of the ray path (black line).  (\emph{Bottom}) The cross-section perpendicular to the ray path.  The intersection
point of the plane and the ray is indicated by the black dot.  The total angular range in both sections is $100^{\mathrm{o}}$, and both
extend radially from 0.3 $R_\sun$ to the surface.} 
\end{figure}

\section{DISCUSSION}

We have presented a different theoretical framework, based on functional analytic tools, that we hope many will find a clear and elegant way
of thinking about traveltime sensitivity kernels.  In addition, we have provided the essential formulas for calculating these kernels 
in a full spherical geometry.  For kernels for medium properties other than the ones discussed here, the mathematical apparatus has been 
developed to allow the reader to derive his or her own formulas.

With some reasonable assumptions about the Sun, we have shown that one can obtain kernels that demonstrate their potential value by 
showing the inadequacy of ray theoretical approximations.  Clearly these results can now be put to good use in analyzing time-distance
measurements to obtain information on the deep structure of the Sun.



\end{document}